\documentclass{llncs}

\usepackage{lipsum}
\usepackage{booktabs}
\usepackage{paralist}
\PassOptionsToPackage{hyphens}{url}
\usepackage{hyperref}
\hypersetup{colorlinks, citecolor=blue, filecolor=blue, linkcolor=blue, urlcolor=blue}
\usepackage{multirow}
\usepackage{subfigure}
\usepackage{graphicx}
\usepackage{fancyhdr}
\usepackage[inline]{enumitem}
\usepackage{amssymb}
\usepackage{chngpage}

\newcommand{\afblock}[1]{\noindent{\textbf{#1}}}
\newcommand{\takeaway}[1]{\noindent{\textbf{Takeaway.}} \textit{#1}}
\newcommand{\sref}[1]{\S\,\ref{#1}}

\usepackage{todonotes}

\begin{document}

\renewcommand{\headrulewidth}{0pt}
\renewcommand{\footrulewidth}{0pt}

\title{DDoS Never Dies? An IXP Perspective on DDoS Amplification Attacks}

\author{Daniel Kopp\inst{1} \and
Christoph Dietzel\inst{1,2}\and\\ 
Oliver Hohlfeld\inst{3}}
\authorrunning{D. Kopp, C. Dietzel, O.Hohlfeld}

\institute{$^{1}$DE-CIX, $^{2}$MPI for Informatics, $^{3}$Brandenburg University of Technology \\
\email{\{daniel.kopp,christoph.dietzel\}@de-cix.net} \\
\email{oliver.hohlfeld@b-tu.de}}

\maketitle
\begin{abstract}
DDoS attacks remain a major security threat to the continuous operation of Internet edge infrastructures, web services, and cloud platforms. While a large body of research focuses on DDoS detection and protection, to date we ultimately failed to eradicate DDoS altogether. Yet, the landscape of DDoS attack mechanisms is even evolving, demanding an updated perspective on DDoS attacks in the wild.
In this paper, we identify up to 2608 DDoS amplification attacks at a single day by analyzing multiple Tbps of traffic flows at a major IXP with a rich ecosystem of different networks. We observe the prevalence of well-known amplification attack protocols (e.g., NTP, CLDAP), which should no longer exist given the established mitigation strategies. 
Nevertheless, they pose the largest fraction on DDoS amplification attacks within our observation and we witness the emergence of DDoS attacks using recently discovered amplification protocols (e.g., OpenVPN, ARMS, Ubiquity Discovery Protocol). 
By analyzing the impact of DDoS on core Internet infrastructure, we show that DDoS can overload backbone-capacity and that filtering approaches in prior work omit 97\% of the attack traffic.
\end{abstract}
\fancypagestyle{firststyle}
{
\fancyfoot[C]{\tiny
Author's edition.
Accepted to Passive and Active Measurement Conference 2021.
Cite as {\em In Proc. PAM 2021}.
}
}
\thispagestyle{firststyle}

\section{Introduction}\label{sec:intro}

With growing relevance for our society and in light of the commercial success of the Internet, naturally also misconduct is increasing. A popular security threat is to launch Distributed Denial-of-Service (DDoS) attacks~\cite{akamaiDDoSreport,github,jonker17ddos} against application or service providers by consuming more critical resources than available, e.g., computing power or network bandwidth. The motivation to conduct in criminal activities are manifold and include financial gain~\cite{ransomware,holz12businessddos}, political motivation~\cite{bank-attack,russian-elections-2012}, and cyber warfare~\cite{estonia2019misc,ukraine2019misc}.

The main reason for the scale of current DDoS attacks~\cite{princesmallddos13,princeddos14,mirai-botnet-usenix-security,2016-Mirai-attack} is the misuse of certain protocols to amplify attack traffic~\cite{akamaiDDoSreport,github,jonker17ddos}. Responses to spoofed traffic~\cite{lichtblau17spoofing,IMC2009Spoofer,BeverlySRUTI2005,Moore2001,spoofer2019}, i.e., packets with modified source IP addresses, are reflected towards the DDoS target and not the original sender. The reflected traffic is not only sent to a different target but also \emph{amplified} as small request can trigger significantly larger responses (up to $\times50,000$)~\cite{uscert18amplification,rossow14amplification,ryba15amplification}. The so-called amplification factor depends on the misused protocol, e.g., NTP, DNS, or more recently Memcached~\cite{rossow14amplification,memcached-Akamai,morales17tbps18}. 

To mitigate these attacks in practice, various reactive DDoS detection and defense techniques filter unwanted traffic of ongoing attacks, e.g., scrubbing services~\cite{Prolexic,Protecting-Websites-COMPUTER,measuring-adoption-ddos,vissers15dns-dps,Scrubber}, blackholing~\cite{Blackholing:PAM2016,measuring-adoption-ddos,dietzel18stellar,BlackholingHoneypots}, or ACLs and Flowspec~\cite{nokia-acl,flowspec}. In this arms race, spontaneously appearing new amplification vectors are quickly growing to cause substantial harm to even well positioned networks and applications~\cite{morales17tbps18,github}. To make matters worse for mitigation service providers and network operators, once exploited protocols for DDoS often remain a threat for decades, despite the joint effort of the research community, operators, and policy makers. While the impact on web services~\cite{sachdeva2009performance,vissers2014ddos} or platform service providers~\cite{sachdeva2009performance} is well studied, only few works study DDoS attacks in the wild. These studies largely rely on measurements taken at the edge at \emph{i)} honeypots~\cite{jonker17ddos,thomas2017ecrime}, \emph{ii)} a DDoS scrubbing service~\cite{Scrubber}, or \emph{iii)} by analyzing network backscatter~\cite{jonker17ddos,DDoSBackscatter}. Only one study analyzes DDoS attacks in Internet traffic captured at the Internet core~\cite{booterIMC} and solely focuses on NTP and Memcached as attack vectors. Thus, a more general study of DDoS attacks visible at the core of the Internet is still missing. Also, while the impact of DDoS attacks on their victims is known, their impact on core Internet infrastructure that forward attack traffic is unknown.

In this paper, we study properties of amplified DDoS attacks in Internet traffic captured at the core of the Internet---at a major Internet Exchange Point. We thereby provide an up-to-date perspective of the current threat landscape and their effects on the IXP itself.
Our major contributions are:
\begin{itemize}[noitemsep,topsep=5pt,leftmargin=9pt]
\vspace{-1em}
\item Well known amplification protocols persist to be the first choice for DDoS attacks and account for 89.9\% of our observed DDoS attacks. Indeed, we find a high number of 14,083 DNS resolvers and 3,637 NTP servers used in attacks.  
\item We provide evidence for the emergence of recently discovered amplification vectors in the wild---with a staggering increase of 500\% within our measurement period---with significant number of reflectors and observed attacks. 
\item We provide insight into the impact of DDoS on infrastructure at the core of the Internet. In general, the IXP and the connected customers were well equipped with sufficient spare capacity.
\item From a view onto targets of DDoS attacks we find networks that received attacks to 28\% of their address space and further find temporal attack patterns.
\item Focusing on a single protocol is not enough: 24\% of the observed victims received DDoS attack traffic using more than one amplification protocol.
\item Port 0 with DDoS attacks can be an artifact of IP fragmentation in flow-traces.
\item By comparing to a commercial world-wide honeypot network, we find largely diverging views: only 8.18\% of the observed attacks (33\% of the target IPs) were also observed by the honeypots.
This provides the first comparison of a core-centric view (here at an IXP) to an edge-centric honeypot perspective that is often used in prior work.
Our results indicate that both perspectives (core Internet and honyepot) have a partial and diverging view.
\end{itemize}

\afblock{Structure.} \sref{sec:data} describes our data set and DDoS detection approach. We study properties of DDoS attacks using new and legacy attack vectors in~\sref{sec:ddos}, their impact on IXP infrastructure in~\sref{sec:ixp}, and their targets in~\sref{sec:targets}. Last, we correlate this new core (IXP) with the traditional edge (honeypot) perspective in~\sref{sec:honeypot}.

\section{Data Sets \& DDoS Classification}\label{sec:data}

\afblock{Data set.}
Anonymized and sampled IPv4 flow-based traffic traces (IPFIX) captured at a major European Internet Exchange Point (IXP) with $>900$ members between Sep.\ 23, 2019 and Apr.\ 20, 2020 with $1.3$T flows were made available to us. They only contain DDoS amplification traffic filtered by our classification scheme and do not contain payload or any further protocol or header information. In addition, the IXP labeled when an  attack was redirected to a connected scrubbing service or if blackholing was enabled for the attacked IP.

\begin{figure}
\centering
	\vspace{-0.3cm}
	\includegraphics[width=0.7\columnwidth]{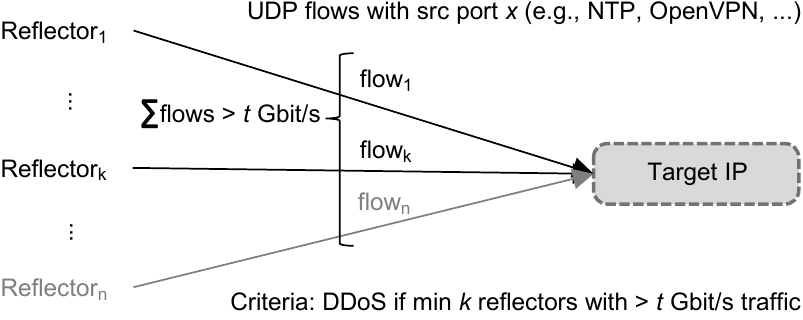}
	\vspace{-0.5cm}
	\caption{We classify traffic as DDoS reflection attack if a target IP gets UDP traffic from at least $k$ sources with an amplification source port and an aggregated rate $>t$\,Gbps.}
	\vspace{-0.5cm}
	\label{fig:detection}
\end{figure}

\afblock{DDoS Classification.}
We use a flow-based classification approach to detect UDP-based DDoS reflection attacks in passive measurement traces as shown in Fig.~\ref{fig:detection}.
We classify traffic as DDoS reflection attack if a target IP receives traffic from at least $k=10$ (total $n \ge k$) IPs with the \emph{same} source port and an aggregate traffic rate of more than $t = 1$ Gbps.
To restrict the filter to servers abused as amplifiers, we require the source port to be a well-known port of UDP-based protocols (e.g., NTP, OpenVPN, DNS) or additionally port 0.
When these criteria match, we refer to the $n$ source IPs as reflectors (i.e., servers sending to the target IP).
We show that typical attacks have a much larger number of reflectors with $n$ being in the order of hundreds or thousands.
Here, the presence of at least $k$ reflectors serves as detection threshold to differentiate DDoS traffic from traditional client-server traffic which could be induced due to legitimate use cases.
In addition, we assume that it is unlikely for a \emph{client} to receive traffic from $k$ sources (\emph{servers}) with the same source port (e.g., NTP time servers) with a high traffic rate $t$.

\afblock{Validation.}
We validated our classification by manually inspecting 300 attack events including all amplification protocols.
With the help of the Internet Exchange Point (IXP) we validate our samples to be plausible cases of DDoS attacks.
The inspection process performed by the IXP included \emph{i)} inspecting customer support cases \emph{ii)} obtaining and examining the traffic levels towards the victim network before, after, and during the potential DDoS attack. All inspected cases where found to be plausible (e.g., victim port traffic levels are atypically high during the attack as compared to other times).
While false positives are still possible, they are unlikely and we did not find cases.
To systematically check for false positives, we examined two widely used protocols: DNS and QUIC.
First, a false positive for DNS would require a target IP to receive more than 1~Gbps of traffic from at least 10 different DNS server IPs. We checked for false positives by high query volumes from authoritative DNS servers from/to a \textbf{root DNS server} collocated at our vantage point and didn't find any.
Last, no \textbf{QUIC} flow---where clients contact a number of web servers---matches our filter criteria.
We cross-check our classification approach for its proneness to false-positives by using QUIC (UDP/443) and alternatively including it into our filter. This approach did not produce any event that matched our classification. We therefore are convinced that our classification process is very well suited for our vantage point. We thus consider all matching flows as DDoS attacks.

\afblock{No impact of COVID-19.}
We remark that the start of the COVID-19 pandemic with global lockdowns and containments falls within our measurement period. While increasing Internet traffic levels were observed during COVID-19 in 2020~\cite{covid19traffic}, we did not observe a noticeable increase in DDoS attacks due to COVID-19 within our measurement time frame.

\afblock{The mysterious case of port 0 as a result of IP fragmentation.}
While reserved~\cite{RFC900} but never assigned and treated as request for a system-allocated port by socket APIs, port 0 should not be observed in Internet traffic.
Prior work~\cite{DNSRootServer,Port0,BOUHARB2014S114,IMC04,dietzel18stellar,OliPort0} observed low volumes of port 0 Internet traffic.
Its origin can be multifold, e.g., as target port for DDoS attacks~\cite{Port0} or scanning~\cite{BOUHARB2014S114} and system fingerprinting~\cite{Port0}.
We also observe traffic carrying port 0, yet with a very different reason: IP fragmentation.
In our case of analyzing IPFIX traces, packets that do not contain a transport protocol header due to fragmentation are assigned src and dst port 0 by the collecting switches.
Similar behavior exists for Netflow V5, V9, and IPFIX export from routers from various vendors~\cite{BrianFragmentation}.
Such traces thus falsely suggest the existence of port~0 traffic in the presence of IP fragmentation and care must be taken in the analysis.
When matching single protocol attacks by time and destination, 43\% of our dataset contains port~0 traffic.
Here, we see a strong correlation of port~0 traffic to DDoS attacks using DNS (in avg. +153\% more traffic), CLDAP (avg. +140\%), and Chargen (avg. +91\%).
Since we cannot reassemble port 0 fragments in the obtained IPFIX data to obtain the true port number, we decided to ignore port~0 and rather report clearly identifiable traffic.
This impacts our results as we \emph{under}estimate \emph{i)} the number of attacks passing the threshold and \emph{ii)} the absolute attack volume. 
We remark that this only impacts the reported absolute values (the previous figures provide an approximation by what factor we underestimate attack volume of DNS, CLDAP, and Chargen), not other results and conclusions. %
\section{DDoS in the Wild}\label{sec:ddos}

\begin{figure}[t]
\centering
\includegraphics[width=.85\linewidth]{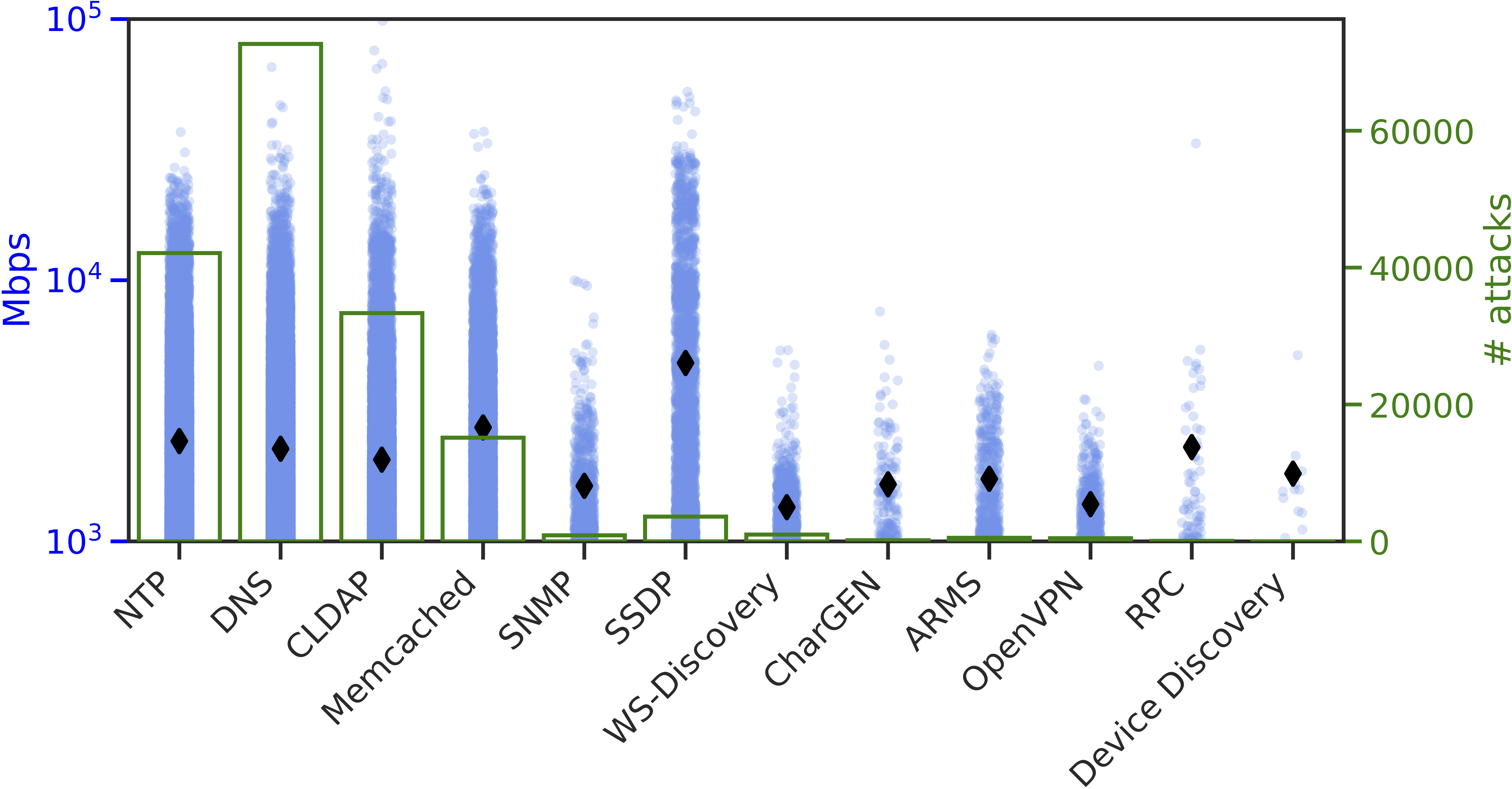}
\caption{Observed DDoS amplification attacks by protocol, with their attack size in Mbps, the median shown as $\blacklozenge$ (left), and the number of attacks per protocol (right).}
\label{fig:ddos_size_distribution}
\vspace{-1em}
\end{figure}

We give a core Internet perspective on the current DDoS attack landscape, beginning by first updating the current state of legacy amplification protocols abused for DDoS attacks.
We then study new protocols that recent DDoS attacks leveraged.
We present details of DDoS attacks we identified according to their amplification protocols in Table~\ref{tab:ddos} and the distribution of their attack volume and frequency in Fig.~\ref{fig:ddos_size_distribution}.
We observe 170,042 events of DDoS attacks which are at least 1~Gbps with the largest one being $98$~Gbps.
Attacks that fall below the 1 $Gbps$ threshold are counted as new event once they exceed 1 $Gbps$ again.
To account for this, we group by day and protocol yielding 97,680 events.
These attacks targeted 58,180 individual IP addresses in 4,433 ASes.
This is 6.5\% of all active ASes and 1.4\% of all advertised prefixes of the Internet.

\begin{table*}[t]
\begin{center}
\scriptsize
\begin{tabular}{lllllllllllllll}
&  & \multicolumn{2}{c}{} & \multicolumn{2}{c}{} & \rotatebox{0}{} & \rotatebox{0}{} & \multicolumn{2}{c}{\textbf{duration}} & \multicolumn{2}{c}{} & \multicolumn{2}{c}{} \\
\cmidrule(l{2pt}r{2pt}){9-10} 
&  &  \multicolumn{2}{c}{\textbf{Gbps}} &  \multicolumn{2}{c}{\textbf{Mpps}} &  &  & $max$ & $avg$ & \multicolumn{2}{c}{\textbf{reflectors}} & \multicolumn{2}{c}{\textbf{pkt size}}  \\
\cmidrule(l{2pt}r{2pt}){3-4} \cmidrule(l{2pt}r{2pt}){5-6} \cmidrule(l{2pt}r{2pt}){11-12} \cmidrule(l{2pt}r{2pt}){13-14} 
\textbf{protocol} & \textbf{port} & $max$ & $avg$ & $max$ & $avg$ & \textbf{targets} & \textbf{attacks} & $days$ & $min$ & $max$ & $avg$ & $avg$ & $std$ &  \\
\midrule
CLDAP & 389 & \textbf{98} & 2.1 & 64.84 & 1.36 & 12,086 & \textbf{33,354} & 3.85 & 6.4 & 2,040 & 328 & 1515 & 21 \\
DNS & 53 & \textbf{66} & 2.3 & 43.48 & 1.54 & 29,023 & \textbf{72,679} & 2.05 & 6.0 & 14,083 & 776 & 1474 & 59 \\
SSDP & 1900 & \textbf{53} & 4.8 & \textbf{150.4} & 13.9 & 1,036 & 3,618 & 7,49 & \textbf{30} & 11,102 & 1594 & 347 & 9.1 \\
Memcached & 11211 & \textbf{37} & 2.7 & 46.87 & 2.51 & 7,119 & \textbf{15,151} & 1,42 & 6.0 & 1,556 &\textbf{35.6} & 1285 & 207 \\
NTP & 123 & \textbf{37} & 2.4 & 77.22 & 5.03 & 21,853 & \textbf{42,124} & 3,04 & 6.5 & 3,637 & 164.7 & 481.1 & 10 \\
RPC & 111 & \textbf{33} & 2.3 & 36.27 & 3.5 & 37 & 73 & 0.02 & 4.7 & 12,217 & 1465 & 620.6 & 51 \\
SNMP & 161 & 9.9 & \textbf{1.6} & 9.32 & 1.21 & 577 & 885 &5.52 & 9.0 & 3,541 & 506 & 1372 &  160 \\
Chargen & 19 & 7.6 & \textbf{1.7} & 6.05 & 1.35 & 105 & 168 & 0.04 & 7.4 & 577 & 247 & 1255 & 145 \\
ARMS & 3283 & 6.2 & \textbf{1.7} & 5.87 & 1.65 & 253 & 519 & 0.18 & 11 & 1,026 & 345 & 1053 & 1.3 \\
WS-Dis. & 3702 & 5.4 & \textbf{1.4} & 5.15 & 1.14 & 485 & 994 & 0.11 & 4.8 & 1,731 & 669 & 1216 & 199 \\
Device Dis. & 10001 & 5.2 & \textbf{1.8} & 24.33 & 8.7 & 10 & 13 & 0.01 & 6.5 & \textbf{7,681} & 2993 & 207.9 & 3.2 \\
OpenVPN & 1194 & 4.7 & \textbf{1.4} & 72.98 & \textbf{21.5} & 385 & 464 & 0.08 & 7.1 & \textbf{8,987} & 3736 & 64.5 & 0.3 \\ \bottomrule
\end{tabular}
\caption{Details about the discovered attacks (size in Gbps and packet rate in Mpps, number of targets, attacks and duration) and observed amplification protocol features (number of reflectors, average packet size (pkt) and their standard deviation in byte).}
\label{tab:ddos}
\end{center}
\end{table*}

\vspace{-0.5em}
\subsection{The State of Legacy DDoS Protocols}

There exist a set of widely studied protocols---e.g., NTP~\cite{czyz2014,rossow14amplification,booterIMC}.
Years have passed since the disclosure of the vulnerability to abuse NTP as amplification vector for DDoS attacks.
The attack is well understood and workarounds or solutions are known for years---in principle, this attack vector should no longer exist.
In 2014, an extensive measurement study~\cite{czyz2014} ``chronicle[s] the rapid rise and steady decline of the NTP DDoS attack phenomenon'', concluding that the operations communities' ``efforts have had a visible impact in diminishing the vulnerable amplifier population and reducing attack traffic''.
Yet, NTP is still a popular vector for DDoS attacks~\cite{booterIMC} and by the rise of further protocols being abused for DDoS the attack landscape continues to increase.
Well known other legacy protocols abused for amplification DDoS are DNS, Chargen, SNMP, and SSDP, whose vulnerability have been known since 2014~\cite{rossow14amplification}.
For some, e.g., DNS, no documented solution exists to generally prevent abuse for DDoS.
We thus focus first on updating the current state of DDoS attacks using legacy protocols.

\textbf{State of legacy amplification protocol attacks today.}
We find CLDAP, NTP, and DNS-based DDoS attacks to still account for 89.9\% of all our observed attacks (Table~\ref{tab:ddos})---despite that the relevance of CLDAP and NTP should have declined long ago.
Given the absence of a solution for DNS, we see most attacks using DNS followed by NTP and CLDAP. 
Legacy protocols account for the highest volume attacks from 33~Gbps (RPC) to 98~Gbps (CLDAP). 
Among these protocols we observe attacks with significantly higher rates of packets per second for SSDP, with a peak of 150.4~Mpps, which is 51\% higher than the next protocol in the list (NTP). 
This makes SSDP-based DDoS attacks more dangerous to any packet processing device, compared to other attack vectors. 
Additionally, for SSDP we experience a very high average duration of 30~minutes from 3,618~attacks towards 1,036~targets, which leads to the assumption that this protocol is used in more sophisticated attacks.
Moreover, although RPC is one of least frequent protocol that we observe, it can still generate large volumes of DDoS attacks, similar to the group of popular DDoS protocols. 
SNMP and Chargen are the least powerful of this group.
Within our observation period they account for 1,053~attacks with sufficient attack traffic to impose a threat for most small to medium sized web services. 

Despite the long time that has passed since the disclosure of these DDoS amplification vectors, they are still the dominant protocols abused for DDoS attacks today.
We thus posit that better approaches for closing these attack vectors are indispensable.

\subsection{New Kids on the Block}

Besides the awareness of legacy protocols being exploited for DDoS attacks, new protocols are being abused additionally.
We next focus on newly abused protocols that have received little (Memcached) to no attention in literature so far to be observed in Internet traffic (``Ubiquiti Device Discovery'', ``WS-Discovery'', ``ARMS'', and ``OpenVPN'').
Among them we notice a steep rise for OpenVPN---first observed as reflection protocol end of 2019~\cite{netscout2020}---growing by more than 500\% in the last month of our observation.

\afblock{Memcached:} In 2018, the widely used database caching system Memcached was found to be vulnerable for amplification attacks with to this date unseen high amplification factors of up to 51,200.
Research confirmed the existence of Memcached-based DDoS attacks in the wild~\cite{10.1145/3278681.3278701,booterIMC,8586810}, as well as white papers by the security industry~\cite{memcacheCF,memcacheAkamai}, and tech news~\cite{memcacheZdnet}.
While the existence is known, their prevalence in Internet traffic hasn't been studied yet.
Today, we still see 8.9\% of all attacks using Memcached as amplification protocol.
Beyond Memcached, we report on the prevalence of DDoS attacks leveraging recently discovered attack vectors:

\afblock{Ubiquiti Device Discovery:} In early 2019, a network device discovery protocol was reported to be used as amplification protocol---with 486k potentially vulnerable devices~\cite{Rapid7Ubiquity}.
While the reported amplification factors are inconsistent (between x4 and x35)~\cite{JimTroutman,Rapid7Ubiquity} we observe an average packet size of 207.9 bytes, which supports the statement of an amplification factor of x4~\cite{JimTroutman}. 
The attacks consist of up to 7,681 reflectors which generate a volume of 5.2 Gbps.
\afblock{WS-Discovery:} In mid 2019, WS-Discovery---a protocol used by an increasing number of IoT devices to discover other UPnP devices within a local network---was reported as amplification protocols.
Reports on the number of publicly exposed systems range from 65k~\cite{netscout2020} to 630k~\cite{zdnetWSD} and the amplification factor from x10 to x500. 
We see almost 1,000 cases which misuse the WS-Discovery service as amplification vector, with an average packet size of 1216 bytes.
The largest attack we recorded was 5.4 Gbps combined from 1,700 reflectors, with the longest attack lasting for 2.64 hours.

\afblock{ARMS:} In June 2019, a protocol used for remote desktop management was reported to be used within DDoS attacks. Around 54,000 potential amplification systems have been discovered at the time~\cite{netscoutArms}.
The amplification factor was reported to be x35.5 with two packets being send, the first 32 bytes, and the second packet with 1034 bytes. From our observation we can report an average packet size of 1052.9 bytes. We have seen 519 DDoS attacks towards 253 victims using the ARMS reflection vector during our measurement period.

\begin{figure}[ht!]
\centering
{
	\subfigure{
		\includegraphics[width=0.75\linewidth]{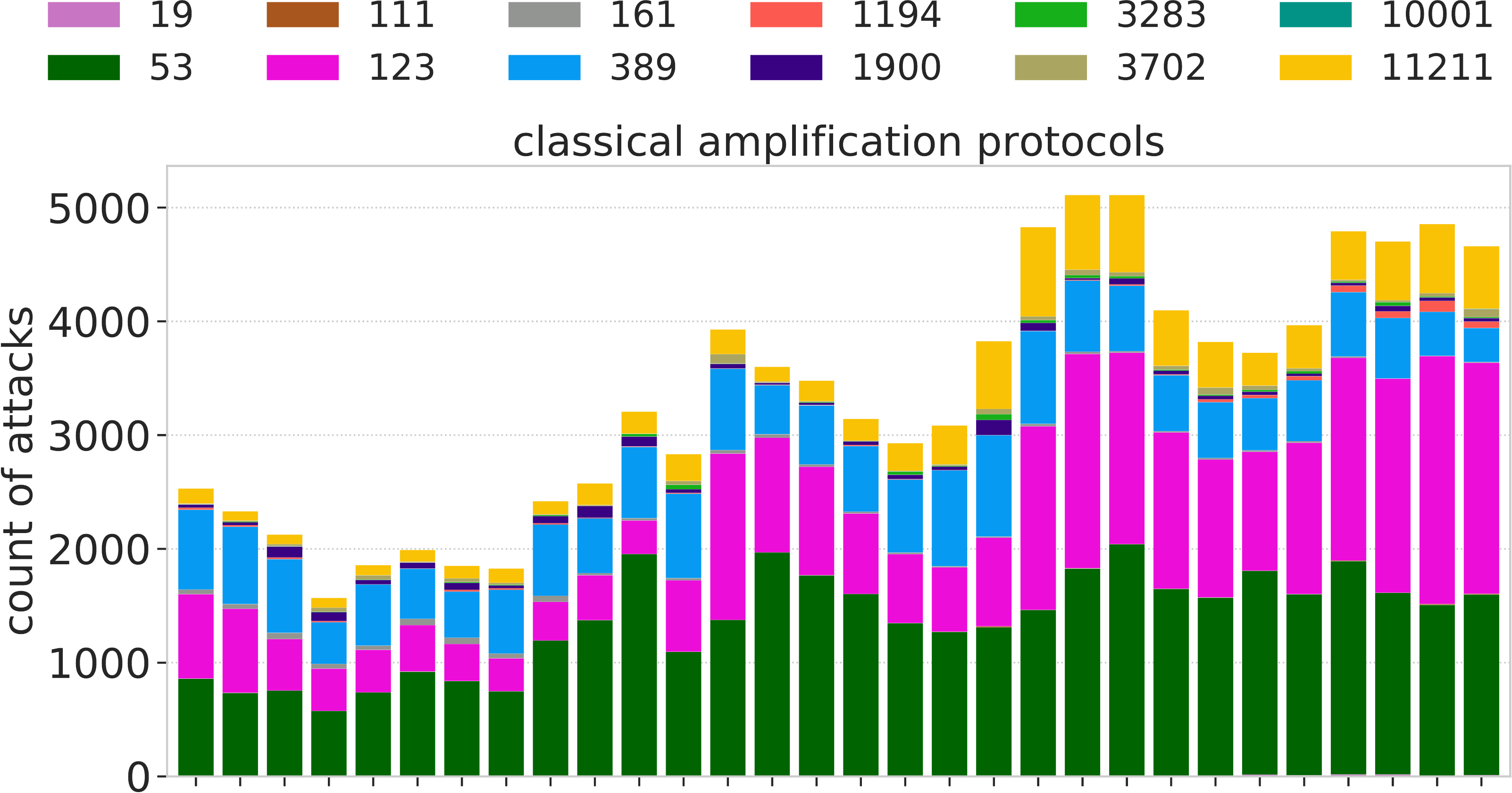}
		\label{fig:ddos1}
	}
	\subfigure{
		\includegraphics[width=0.75\linewidth]{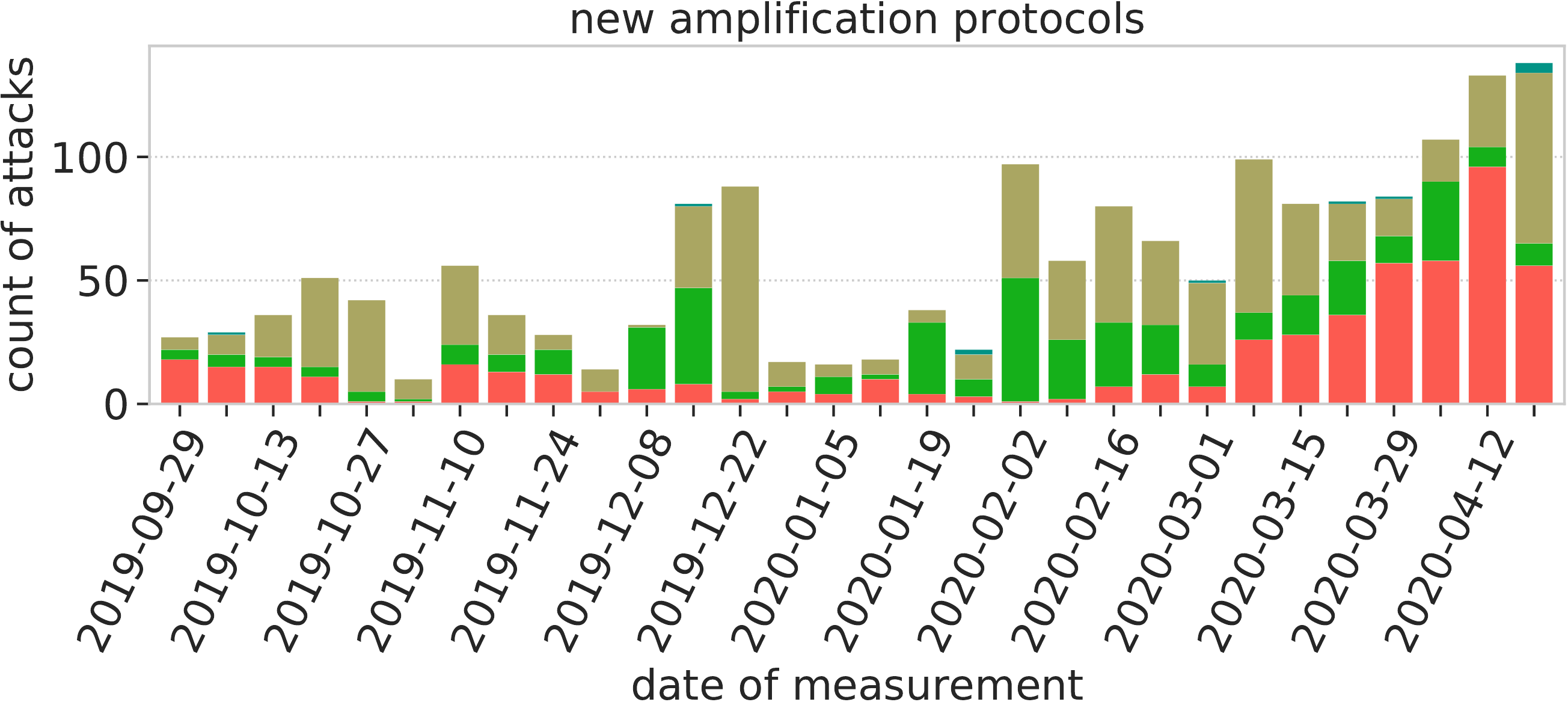}
		\label{fig:ddos2}
	}
	\vspace{-1em}
	\caption{Number of attacks using classical (upper) and new (bottom plot) amplification protocols over time. Bar colors indicate protocol ports and are shared with Fig.~\ref{fig:ddos_pps_mbps}.}
	\label{fig:ddos}
}
\end{figure}

\afblock{OpenVPN:} 
An industry report from 2020 considers OpenVPN as a new attack vector for DDoS attacks~\cite{netscout2020}. An article describes the attack in Sep.\,2019~\cite{FreeBuff} with different vulnerability for reflection attacks allowing for x5 or x60 amplification by replying with multiple packets from one initial packet being send towards the reflector.
We see this attack vector being used by 464 attacks towards 385 targets and up to 2993 reflectors.
We observe an average packets size of 64.5 bytes, supporting the findings of the latest vulnerability report~\cite{FreeBuff}.
Fig.~\ref{fig:ddos} shows an uprise of DDoS attacks within the last month of our study by 500\%.

\takeaway{Beyond anecdotal evidence, we confirm that recently discovered attack vectors in the form of new protocols are being actively abused for DDoS attacks.
Our study quantifies their existence in the wild for the first time.}\\
We shared our findings with an international cyber security technology company with CERT services.
The company is aware of most of the new amplification protocols, but didn't expect them to be already used in the wild. They acknowledged that Table~\ref{tab:ddos} provides a good indication on which attack vectors to include in their mitigation and monitoring solution.
\vspace{-0.25em}
\subsection{Multi-Protocol Attacks}%
It is not enough to focus on just one of the most prominent or upcoming protocols.
Within our dataset we observe 24\% of victims received DDoS attacks by more than one amplification protocol, whereas 4.5\% targets have been attacked with more than two amplification protocols over time.
By investigating few booter services websites (i.e., DDoS as a service platforms see e.g.,~\cite{booterIMC}) and their advertisement, we noticed that new attack methods are being added that are called "MIXAMP" or "ALLAMP"---suggesting the use of all supported amplification protocols to launch attacks.

\subsection{Attack Packet Rates vs.\ Volume}
\begin{figure}
\centering
\vspace{-2em}
\includegraphics[width=0.95\linewidth]{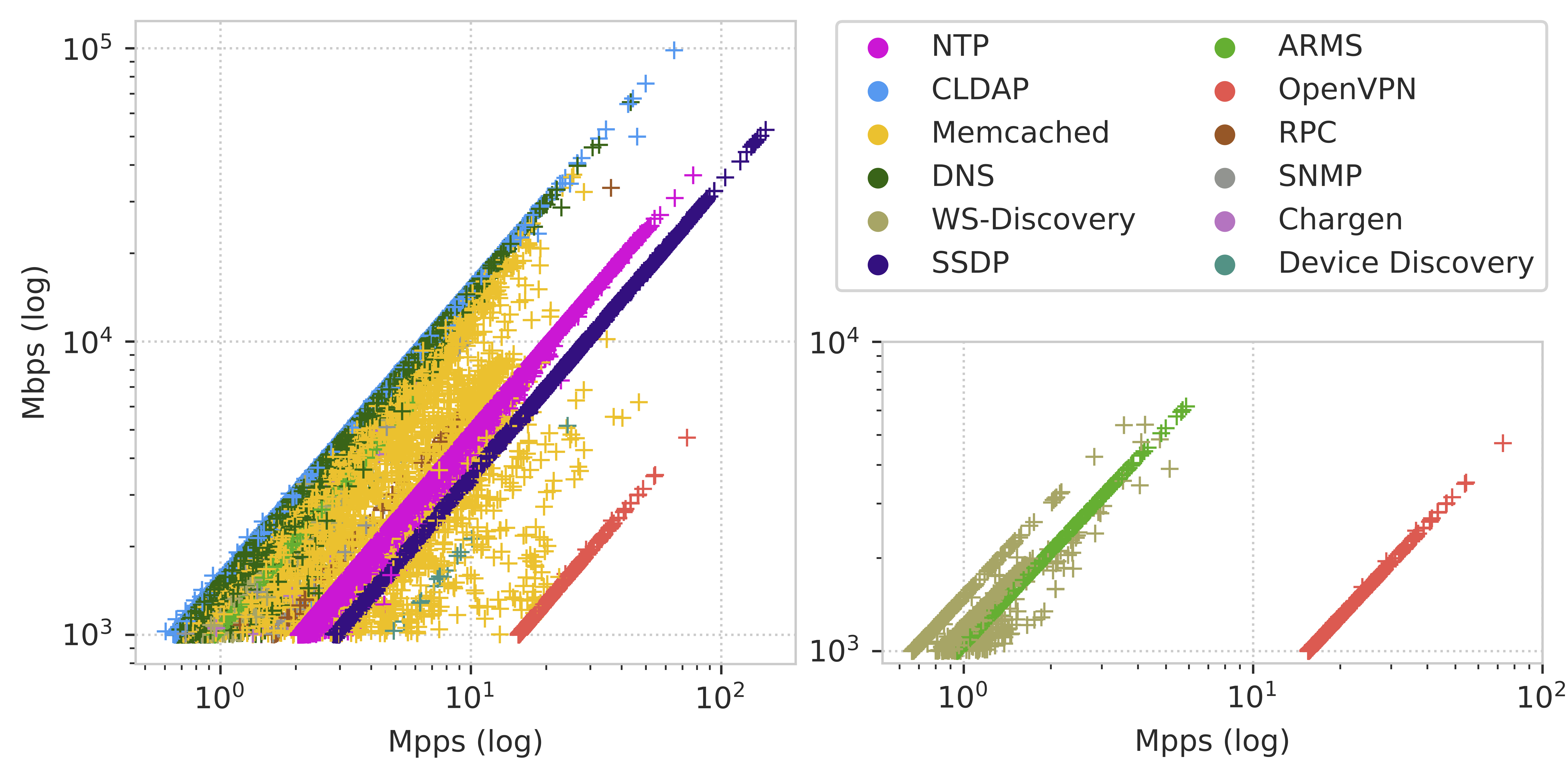}
\vspace{-1em}
\caption{Correlation of packet rate to volume (color set shared with Fig.~\ref{fig:ddos})}
\vspace{-1.2em}
\label{fig:ddos_pps_mbps}
\end{figure}

When the amplification factor is constant, the attack volume can be scaled by the packet rate sent to reflectors.
We thus show the relationship of packet rate and volume for all protocols in Fig.~\ref{fig:ddos_pps_mbps}.
We observe 3 different characteristics:

\afblock{Single linear relationships.}
For most attacks, we observe a linear relationship between the packet rate and the attack volume size, hinting to a constant amplification factor.
This is visible as straight lines in Fig.~\ref{fig:ddos_pps_mbps} (e.g., OpenVPN on the right-hand side of the figure).
We confirmed this relationship for every protocol by fitting linear regression models (not shown).
There are, however, two protocols that diverge from this simple linear relationship that we describe next.

\afblock{Multiple linear relationships.}
In the case of WS-Discovery we observe multiple linear relationships.
These are indicated in Fig.~\ref{fig:ddos_pps_mbps} in the lower plot at the right.
This indicates that different protocol features are exploited for the attack, each yielding a different amplification factor.

\afblock{No observable relationship.}
Memcached amplification is not linear in terms of packets to volume output, we observe a great variance of the packet rate to Mbps ratio. This effect can have two reasons, either Memcached behaves unpredictably for attackers due to variable response sizes and thus amplification factors, or the response of the Memcached server is controlled by the attacker to insert records retrieved for the attack.

\afblock{Observed volumes.}
DNS and CLDAP provide the highest volumetric DDoS attacks, OpenVPN on the other end is able to generate significant rates of packets while at the same time keeping the traffic volume low.
This means that the highest volumetric attack we observed, with 98 Gbps, had a rate of 64.84 Mpps, whereas OpenVPN recorded a higher rate of packets with 72.98 Mpps and just 4.7 Gbps of volume. Nevertheless, the highest packet rate during our measurement period was due to a SSDP attack with 150.4 Mpps and 53 Gbps. 
Both ends of this scale (CLDAP and OpenVPN) can be favorable to attackers, as they either might want to maximize their invest on sent packets in terms of attack volume or they might want to be as stealth as possible regarding volume but maximizing the impact on packet processing devices.

\afblock{Theoretical maximum volume.}
The results presented above raise the question on how large a combined DDoS attack can become. Assuming one could use all reflectors observed in one week of our measurements, we estimate a DDoS attack with at least 0.875 Tbps to be practical feasible. We use the average output Mbps per reflector, that we calculate from Table~\ref{tab:ddos} and multiply with number of unique amplifiers that we can observe over the course of one week.
\section{Infrastructure Perspective}\label{sec:ixp}
We use the unique perspective of an IXP as infrastructure provider carrying traffic of more than 900 different ASes, and therefore also hundreds of substantial DDoS attacks.
In particular, the challenge is to withstand the combined volume of many DDoS attacks simultaneously. 
In this section, we provide an infrastructure perspective on DDoS attacks.
\afblock{IXP Infrastructure.}
At the measured IXP, the highest share of attack traffic forwarded due to multiple parallel DDoS events is 3.16\% of the highest daily maximum traffic volume.
The transported attack traffic is only a small share compared to the legitimate traffic and we find no evidence for DDoS traffic to impact the IXP's infrastructure.
In theory, backbone capacity of infrastructures like IXPs cannot be overwhelmed by volumetric DDoS attacks due to the basic nature of their topology: the ingress equals the egress capacity.
In reality, core Internet infrastructures are evolving and becoming more complex, conserving bandwidth over connections between locations and leased fibers is of growing economic interest~\cite{GlobePEER}.

\begin{figure}
\centering
\vspace{-2em}
\includegraphics[width=.7\linewidth]{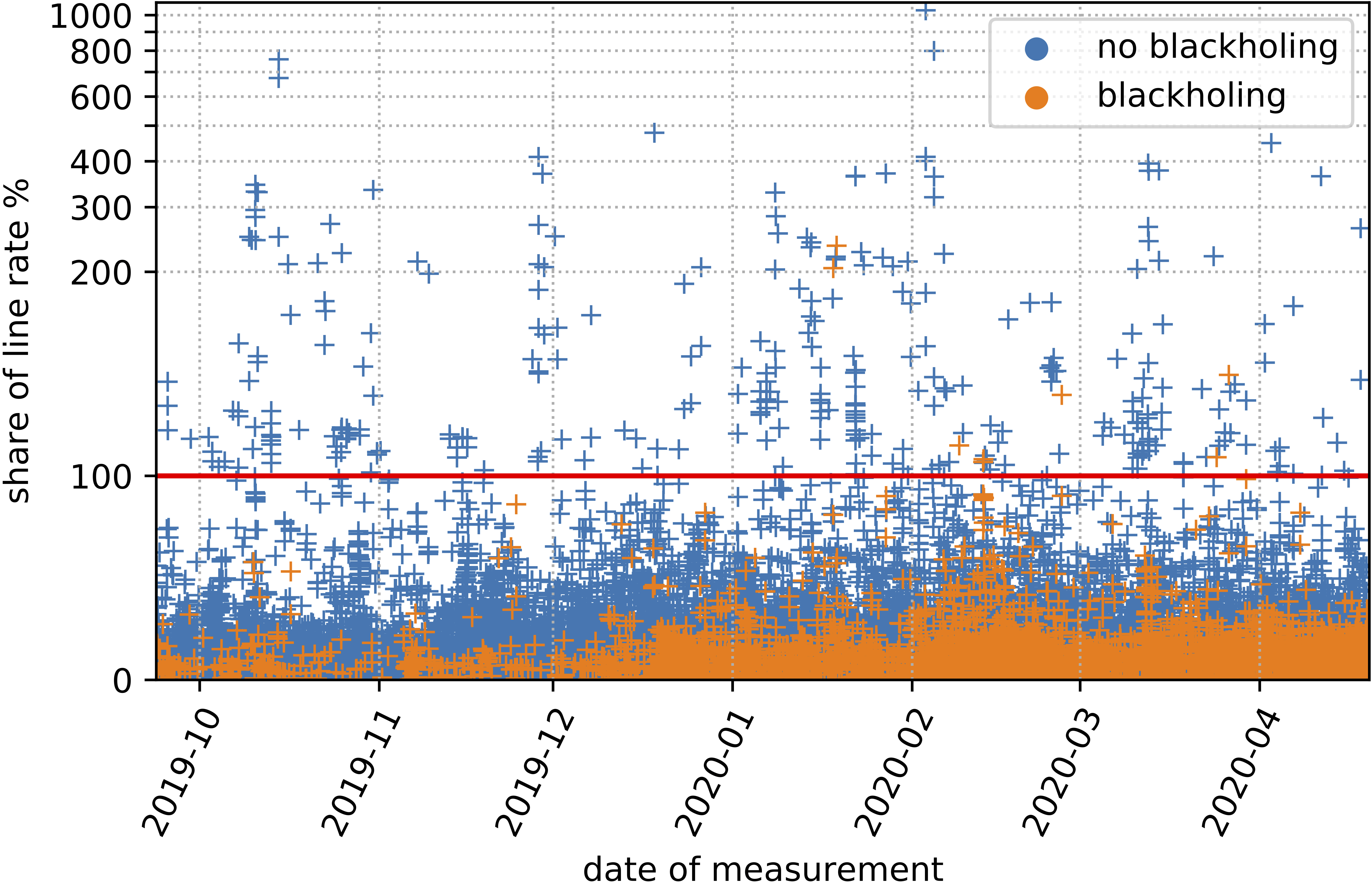}
\vspace{-1em}
\caption{Link capacity in relation to attack size.}
\vspace{-1.5em}
\label{fig:ddos_linerate}
\end{figure}

\afblock{IXP Ports.}
We study the DDoS volume in relation to the port capacity towards the victim's infrastructure (i.e., backbone links to other networks) for all 170k attacks and show it in Fig.~\ref{fig:ddos_linerate}.
The maximum port capacity is indicated by a red horizontal line at 100\%.
Notably, we find 306 cases (0.18\% of all attacks towards 48 individual networks) where DDoS attack surpassed the available capacity of the links at the IXP.
We remark that traffic $>100\%$ of an egress port's capacity can traverse the peering platform from many other members at the IXP and arrive at the given port leading to packet loss.
Of these 48 networks, 12\% had a port capacity below 2 Gbps and 82\% more than 10 Gbps. 
The average duration of this group of attacks is 21 minutes, which shows that these cases are not short bursts, but attacks that overwhelmed the port capacity for a notable time.  
We learn that our observed DDoS attacks are rarely larger than the size of the IXP member's egress port capacity.
This view, however, ignores the typical utilization of the port.
DDoS attack that require up to 50\% of the network's egress link are seen for 26\% of the attacked networks and this additional port utilization might already have led to packet loss and collateral damage at the target network.

\afblock{DDoS Mitigation.}
To mitigate attacks, networks providers have two main tools available. 
One is to contract a DDOS mitigation services to scrub DDoS traffic and forward only legitimate traffic.
Another option is to discard traffic for specific prefixes at the IXP before reaching their target network by using so-called blackholing (labeled in our data).
The deployment of blackholing has been studied widely by previous work~\cite{mskix15,netix15,Blackholing:PAM2016,Giotsas17-blackholing,dietzel18stellar,downTheBlackhole} which focuses on analyzing characteristics of the blackholed traffic and the activation of blackholing events. We observe that only 3.82\% of the DDoS attacks in our dataset are \emph{blackholed}, i.e., the victim asked the IXP to discard traffic to the attacked IP by a blackholing announcement in BGP (labeled by the IXP in our data set).
In as few as 145 cases we saw a redirection of traffic towards to an external \emph{scrubbing} service directly connected to the IXP.
Thus, if blackholing is used as classifier to study attacks in prior work~\cite{downTheBlackhole}, the bulk of the DDoS traffic in our data set is omitted.

Next, we analyze DDoS amplification vectors that are mitigated by blackholing in comparison to all used DDoS amplification vectors.
Whereas NTP attacks in the wild are only the second most prominent attack vector with 24.77\%, they are mitigated the most with 58\% of all blackholing events. 
The most prominent attack vector we observed, DNS with 42.74\%, only has a share of 16.89\% within mitigation. Memcached is mitigated with a share of 15.65\% (in the wild 8.91\%) and CLDAP with 8.31\% (19.62\% in the wild). This reveals a shift of NTP, and Memcached attacks being mitigated more frequently compared to DNS and CLDAP attacks relative to their occurrence. In Fig.~\ref{fig:ddos_linerate}, we see that 63\% of the blackhole events correlate to DDoS traffic lower than 10\% of the networks port capacity. In only 1.1\% of the events the DDoS traffic was $>50\%$ of the capacity. 

Looking at the delay from the start of the attack to the deployment of a mitigation, we see an average delay of 1.16 minutes for the blackholing. 70\% of blackholing rules were installed prior to when we first detect the DDoS attack. 
In addition, we see a delay of $<$10 min for 98.7\% and a delay of $>$4 min in 4.2\% of all blackholing deployments. Only in 19 cases we record a delay greater than 30 minutes, with the highest delay being 5 hours for an 8-hour long attack. These findings are similar to prior work~\cite{BlackholingHoneypots}, that describes a delay of $<$10 min for 84.2\% within their data set. The low attack volume in relation to the port capacity of blackholing events, in combination with the short delay, suggest an automation of the blackholing mitigation.

\takeaway{While the share of DDoS traffic at the IXPs overall infrastructure is insignificant, it can exceed the port capacity of individual customers and thereby impact legitimate traffic.
Blackholing as a DDoS defense technique was used in only 3\% of the attacks we observed and therefor cannot be reliably be used as the sole criterion to report on the state of DDoS in the wild.}

\section{View on Targets}\label{sec:targets}

Last, we analyze the victims of the observed DDoS attacks. 
We study how the DDoS attack landscape is distributed over different networks types and services.

\afblock{Network types.}
We aggregate victim networks by their infrastructure type according to PeeringDB. 
While the average attack volume is mostly the same for each class, some classes are attacked more frequently.
While non-profit networks receive the least amount of attacks (0.06\%), content hosting networks were attacked the most with 36.97\% of all DDoS attacks.
Enterprise and the remaining classes have a comparably low share on the attacks in our dataset. 
Beyond content, eyeball networks (cable/DSL/ISP and NSP) also receive a large number of attacks (34.51\%) that we can attribute to residential users.
This is in line with prior work showing that booter-based DDoS attacks are often launched by online-gamers against other players~\cite{karami2013booters}.

\afblock{Share of attacked address space.}
To understand if any targeted attacks against specific organizations exist, we study the share of attacked address space of individual networks. 
Most significantly, we observe a US based cloud payroll provider where DDoS attacks targeted 28\% of the AS's IP space.
With 16\% a small network of a state bank in the south east Mediterranean region has been the victim of DDoS attacks.
Furthermore, we see attacks that account for 15\% address space of a south Korean cloud provider, and 10\% of an US insurance with 19 hours combined attack time.

\afblock{DNS.}
To better understand the attacked infrastructure, we match the victim IPs to weekly DNS resolutions for \texttt{www.} labels, NS, and MX records of 200M+ domain names obtained from DNS zone files (including .com/net/org and new gTLDs)~\cite{netrayPoster} during our measurement period.
We can match 94.3\% of the attacked IPs to DNS records. 
For 58.63\% we find a matching \texttt{www.} label, suggesting the target to be a web server.
For 27.23\% we find a matching mail exchange (MX) and for 14.14\% a matching authoritative DNS server (NS).
\afblock{VPN.}
VPN service are a relevant service that enables remote work, e.g., during COVID-19 lockdowns.
To find attacks against VPN services, we identify IPs labeled as \texttt{*vpn*} but not as \texttt{www.} in the DNS by searching for *vpn* in any domain label left of the public suffix (e.g., \texttt{companyvpn3.example.com}) in \emph{i)} 2.7B domains from TLS in CT Logs from 2015---2020, \emph{ii)} 1.9B domains from Rapid7 resolutions of reverse DNS, zonefiles, TLS certificates of March, and \emph{iii)} 8M domains from the Cisco Umbrella top list in 2020. 
This gives us 1,2M unique VPN IPs. 
However, we only observed 101 attacks against 39 IPs in 30 ASes and no noticeable increase in the last months.
This attack vector is (fortunately) not yet widely exploited.
Despite, we posit that enterprises should consider protecting their VPNs from DDoS before widespread attacks emerge.

\begin{figure}[!htb]
    \centering
    \begin{minipage}{.5\textwidth}
        \centering
        \includegraphics[width=1\linewidth]{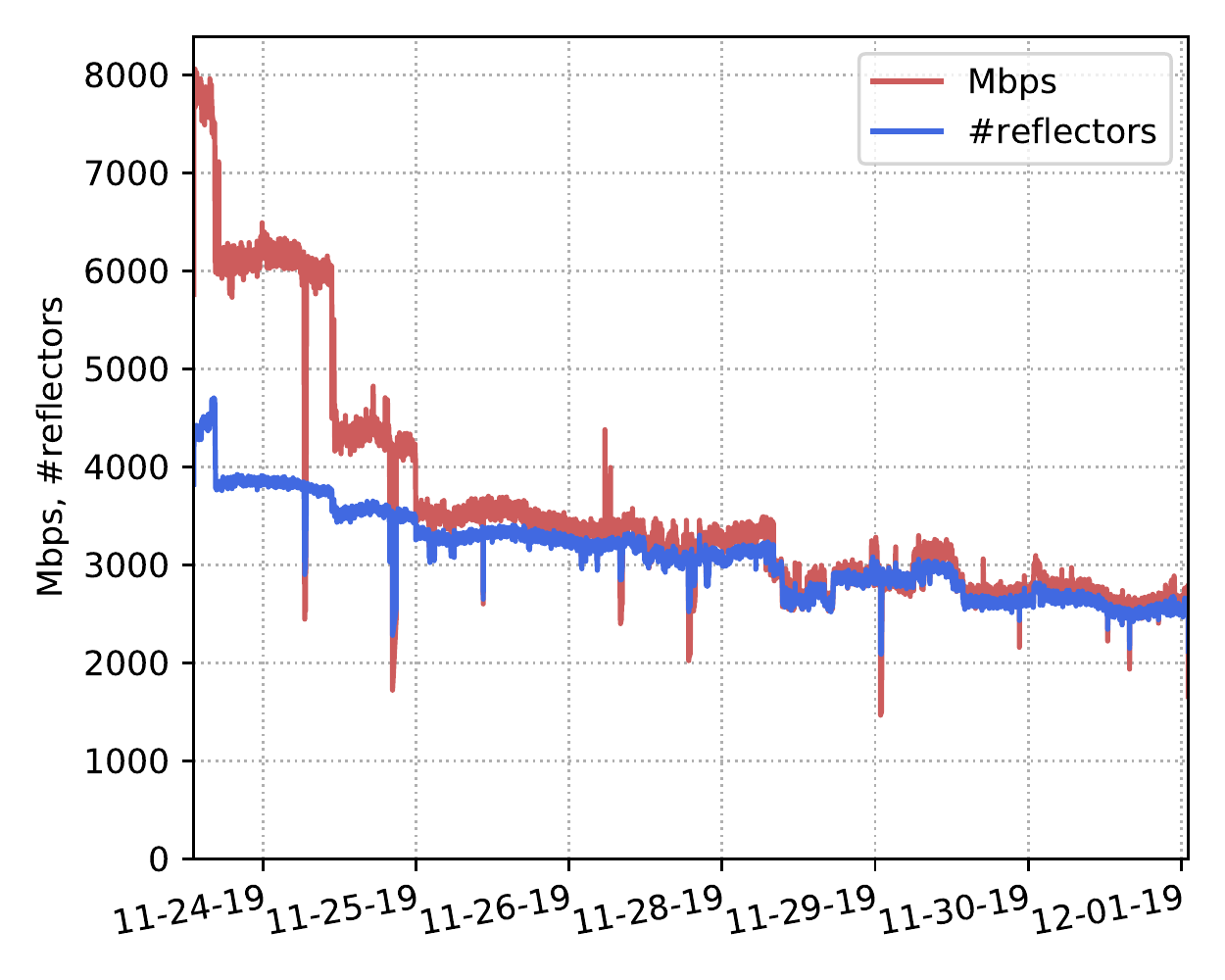}
	\vspace{-3em}
        \caption{SSDP attack over 7 days.}
        \label{fig:longest_attack}
    \end{minipage}%
    \begin{minipage}{0.5\textwidth}
        \centering
        \includegraphics[width=1\linewidth]{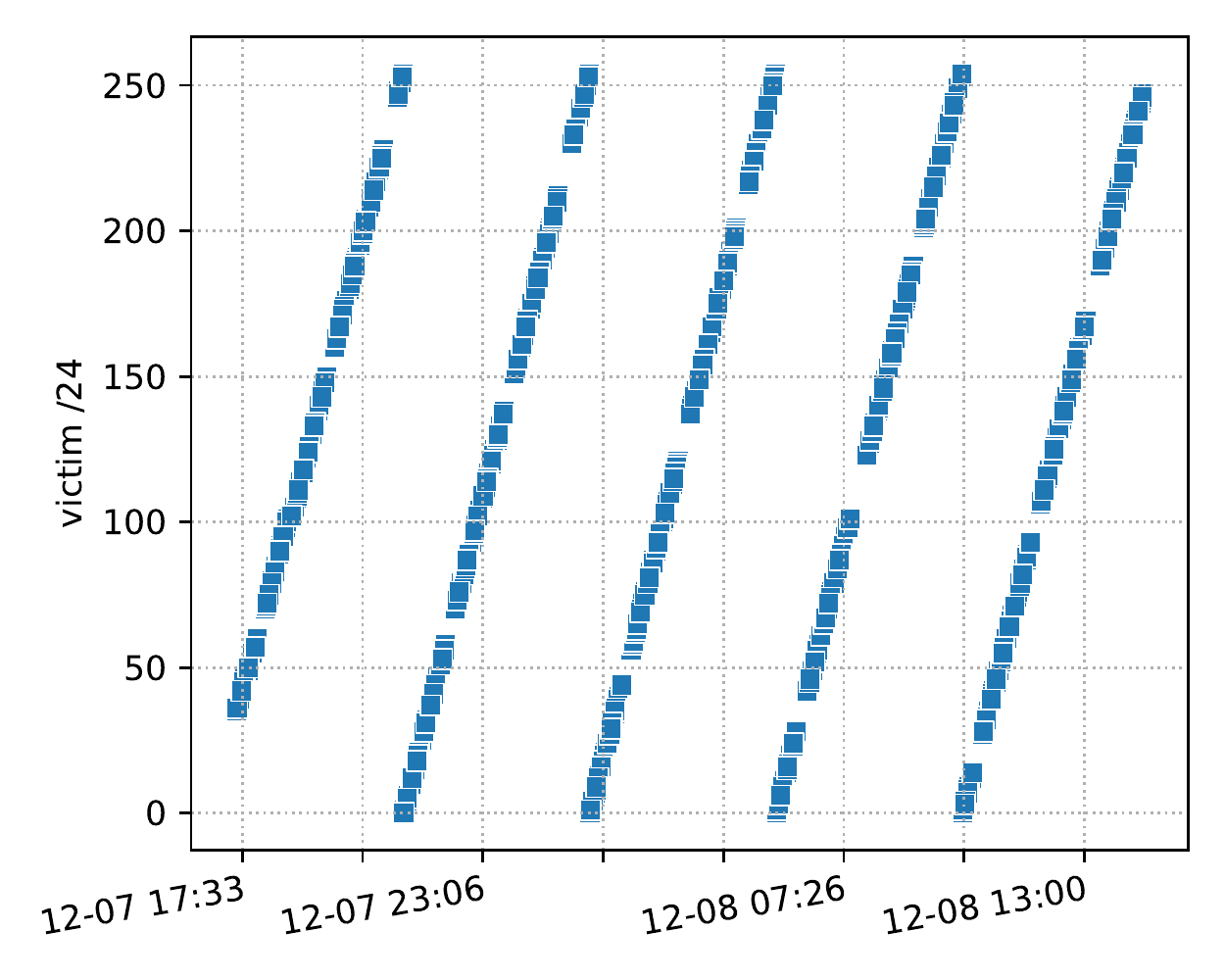}
	\vspace{-3em}
        \caption{DDoS onto /24 network.}
        \label{fig:loop_attack}
    \end{minipage}
	\vspace{-1em}
\end{figure}

\afblock{Temporal DDoS attack pattern.}
We report on two notably cases of DDoS attack. Fig.~\ref{fig:longest_attack} shows the longest consecutive attack within our study. The attack used the SSDP protocol and lasted for 7 1/2 days, with a peak at 8 Gbps and 23.5 Mpps. The attack was targeted against a Swedish Broadband network, whose backbone link never fully saturated. Second, we find a case of a DDoS attack against a Ukrainian ISP (Fig.~\ref{fig:loop_attack}) using DNS as attack vector, attacking one IP address every 1 minute by consecutively traversing a /24 network range. We found other similar temporal attack patterns, where the attack changed between two IPs every minute within an attacked network address space.

\takeaway{By focusing at the victims of amplification DDoS attacks, we find content and eyeball networks to be the most prominent targets. Due to the importance of being able to work from home during the COVID-19 outbreak, we take a look onto attacks towards VPN infrastructures, where we observe 101 attacks against 39 victims. Finally, we observe an interesting attack pattern, where an attacker changed the target IP within a victim’s networks every minute, potentially to evade DDoS mitigation.}

\section{Honeypot Perspective}\label{sec:honeypot}
Honeypots are a widely used tool to study DDoS attacks (see e.g.,~\cite{kramer2015raid,jonker17ddos,thomas2017ecrime}), e.g., setup at universities as single vantage point.
To put our measurements into perspective, we obtained DDoS attacks observed by the world-wide distributed honeypot network operated by CrowdStrike matching our measurement period.
The dataset contains 3.3M events.
We find largely diverging views: only 8.18\% of the observed attacks (33\% of the target IPs) were also observed by the honeypots.
The missing 67\% of targets in the honeypot dataset can be explained by the low likelihood of an attack choosing the honeypots as reflectors.
In turn, our dataset only represents 0.95\% of the targets visible in the honeypot dataset, this is likely due to our robust classification criterion of attacks being $<1$ Gbps, which misses any attack with a lower volume at the IXP. Other factors for the honeypot containing events and targets missing within our dataset are the limited view of the IXP onto Internet traffic and the different location within the Internet's topology.
In contrast the honeypot dataset also consists of low volume and scanning events.
Also the attack protocol popularity diverges, highlighted by 58\% of the honyepot captured events to be Memcached.

\takeaway{
Our results put the use of honeypots (edge measurements, typically used in the literature) into a core Internet perspective and indicate that both perspectives (core Internet and honyepot) have a partial and diverging view.
We thus posit that future research should take multiple perspectives to obtain a more complete view on the DDoS threat.
}
\section{Conclusions}\label{sec:concl}

This paper provides an updated perspective on the state of DDoS amplification attacks and protocols in the wild.
Despite the prediction and hope that the relevance of long-known legacy amplification protocols (e.g., NTP) will decline, we show that opposite is true: CLDAP, NTP, and DNS-based DDoS attacks account for 89.9\% of all observed attacks. In addition, we show that recently disclosed amplification protocols are already used to perform DDoS attacks and can generate effective attacks, e.g., for OpenVPN we even record a 500\% rise within our measurement period.
By taking a view onto the infrastructure at the core of the Internet, we see no severe impact due or degradation of network quality.
We further show that honeypots---typically used to study DDoS---can provide a different picture than the one by traffic captures at the Internet core.
\section*{Acknowledgments}

We thank the anonymous reviewers and our shepherd, Amogh Dhamdhere (Amazon Web Services), for
their constructive comments. We further thank Mark Schloesser and CrowdStrike for their comments and for providing honeypot data.
This work was supported by the German Federal Ministry of Education and Research (BMBF) project AIDOS (16KIS0975K, 16KIS0976). 

\bibliographystyle{splncs04}
\bibliography{references}

\begin{thebibliography}{10}
\providecommand{\url}[1]{\texttt{#1}}
\providecommand{\urlprefix}{URL }
\providecommand{\doi}[1]{https://doi.org/#1}

\bibitem{JimTroutman}
{Jim Troutman via Twitter}.
  \url{https://twitter.com/troutman/status/1090212243197870081}, accessed:
  2020-05-26

\bibitem{Prolexic}
Akamai: {Prolexic Technologies by Akamai}.
  \url{https://www.akamai.com/us/en/cloud-security.jsp} (2018)

\bibitem{memcached-Akamai}
Akamai: {State of the Internet Security Report (Attack Spotlight: Memcached)}.
  \url{https://www.akamai.com/us/en/multimedia/documents/state-of-the-internet/soti-summer-2018-attack-spotlight.pdf}
  (2018)

\bibitem{memcacheAkamai}
Alerts, A.S.: Memcached-fueled 1.3 tbps attacks.
  \url{https://blogs.akamai.com/2018/03/memcached-fueled-13-tbps-attacks.html}
  (2018)

\bibitem{mirai-botnet-usenix-security}
Antonakakis, M., April, T., Bailey, M., Bernhard, M., Bursztein, E., Cochran,
  J., Durumeric, Z., Halderman, J.A., Invernizzi, L., Kallitsis, M., Kumar, D.,
  Lever, C., Ma, Z., Mason, J., Menscher, D., Seaman, C., Sullivan, N., Thomas,
  K., Zhou, Y.: {Understanding the Mirai Botnet}. USENIX Security Symposium
  (2017)

\bibitem{russian-elections-2012}
{BBC}: {'Hacking attacks' hit Russian political sites}.
  \url{http://www.bbc.com/news/technology-16032402} (2012)

\bibitem{IMC2009Spoofer}
Beverly, R., Berger, A., Hyun, Y., claffy, k.: {Understanding the Efficacy of
  Deployed Internet Source Address Validation Filtering}. In: ACM IMC (2009)

\bibitem{BeverlySRUTI2005}
Beverly, R., Bauer, S.: The spoofer project: Inferring the extent of internet
  source address filtering on the internet. In: Steps to Reducing Unwanted
  Traffic on the Internet Workshop (2005)

\bibitem{netscoutArms}
Bjarnason, S., Dobbins, R.: {Netscout - A Call to ARMS: Apple Remote Management
  Service UDP Reflection/Amplification DDoS Attacks}.
  \url{de.netscout.com/blog/asert/call-arms-apple-remote-management-service-udp}
  (2020)

\bibitem{DDoSBackscatter}
Blenn, N., Ghi\"{e}tte, V., Doerr, C.: Quantifying the spectrum of
  denial-of-service attacks through internet backscatter. In: International
  Conference on Availability, Reliability and Security (2017)

\bibitem{BOUHARB2014S114}
Bou-Harb, E., Lakhdari, N.E., Binsalleeh, H., Debbabi, M.: Multidimensional
  investigation of source port 0 probing. Digital Investigation  \textbf{11},
  114 -- 123 (2014)

\bibitem{DNSRootServer}
Brownlee, N., kc~Claffy, Nemeth, E.: {DNS measurements at a root server}. In:
  IEEE GLOBECOM (2001)

\bibitem{10.1145/3278681.3278701}
Burke, I.D., Herbert, A., Mooi, R.: {Using Network Flow Data to Analyse
  Distributed Reflection Denial of Service (DRDoS) Attacks, as Observed on the
  South African National Research and Education Network (SANReN): A Postmortem
  Analysis of the Memcached Attack on the SANReN}. In: Annual Conference of the
  South African Institute of Computer Scientists and Information Technologists
  (2018)

\bibitem{holz12businessddos}
B{\"u}scher, A., Holz, T.: {Tracking DDoS Attacks: Insights into the Business
  of Disrupting the Web}. In: USENIX Workshop on Large-Scale Exploits and
  Emergent Threats (2012)

\bibitem{zdnetWSD}
Cimpanu, C.: {ZDNet - Protocol used by 630,000 devices can be abused for
  devastating DDoS attacks}.
  \url{www.zdnet.com/article/protocol-used-by-630000-devices-can-be-abused-for-devastating-ddos-attacks/},
  accessed: 2020-05-26

\bibitem{flowspec}
{Cisco}: {Implementing BGP Flowspec}.
  \url{https://www.cisco.com/c/en/us/td/docs/routers/asr9000/software/asr9k_r5-2/routing/configuration/guide/b_routing_cg52xasr9k/b_routing_cg52xasr9k_chapter_011.html}
  (2018)

\bibitem{memcacheCF}
Cloudflare: {Memcached DDoS Attack}.
  \url{https://www.cloudflare.com/learning/ddos/memcached-ddos-attack/}

\bibitem{czyz2014}
Czyz, J., Kallitsis, M., Gharaibeh, M., Papadopoulos, C., Bailey, M., Karir,
  M.: {Taming the 800 Pound Gorilla: The Rise and Decline of NTP DDoS Attacks}.
  In: ACM IMC (2014)

\bibitem{GlobePEER}
{DE-CIX}: {DE-CIX GlobePEER Remote}.
  \url{https://www.de-cix.net/de/de-cix-service-world/globepeer-remote} (2018)

\bibitem{Blackholing:PAM2016}
Dietzel, C., Feldmann, A., King, T.: {Blackholing at IXPs: On the Effectiveness
  of DDoS Mitigation in the Wild}. PAM  (2016)

\bibitem{dietzel18stellar}
Dietzel, C., Wichtlhuber, M., Smaragdakis, G., Feldmann, A.: {Stellar: Network
  Attack Mitigation using Advanced Blackholing}. In: ACM CoNEXT (2018)

\bibitem{covid19traffic}
Feldmann, A., Gasser, O., Lichtblau, F., Pujol, E., Poese, I., Dietzel, C.,
  Wagner, D., Wichtlhuber, M., Tapiador, J., Vallina-Rodriguez, N., Hohlfeld,
  O., Smaragdakis, G.: {The Lockdown Effect: Implications of the {COVID-19}
  Pandemic on Internet Traffic}. In: ACM IMC (2020)

\bibitem{Protecting-Websites-COMPUTER}
Gillman, D., Lin, Y., Maggs, B., Sitaraman, R.K.: {Protecting Websites from
  Attack with Secure Delivery Networks}. IEEE Computer Magazine  \textbf{48-4}
  (2015)

\bibitem{Giotsas17-blackholing}
Giotsas, V., Smaragdakis, G., Dietzel, C., Richter, P., Feldmann, A., Berger,
  A.: {Inferring BGP Blackholing Activity in the Internet}. In: ACM IMC (2017)

\bibitem{Rapid7Ubiquity}
Hart, J.: {Rapid7 - Understanding Ubiquiti Discovery Service Exposures}.
  \url{blog.rapid7.com/2019/02/01/ubiquiti-discovery-service-exposures/},
  accessed: 2020-05-26

\bibitem{netrayPoster}
Hohlfeld, O.: Operating a {DNS}-based active internet observatory. In: ACM
  SIGCOMM Poster (2018)

\bibitem{ukraine2019misc}
Interfax-Ukraine: {Poroshenko reports on DDoS-attacks on Ukrainian CEC from
  Russia on Feb. 24-25}.
  \url{https://www.kyivpost.com/ukraine-politics/poroshenko-reports-on-ddos-attacks-on-ukrainian-cec-from-russia-on-feb-24-25.html}
  (2019)

\bibitem{jonker17ddos}
Jonker, M., King, A., Krupp, J., Rossow, C., Sperotto, A., Dainotti, A.:
  {Millions of targets under attack: a macroscopic characterization of the DoS
  ecosystem}. In: ACM IMC (2017)

\bibitem{measuring-adoption-ddos}
Jonker, M., Sperotto, A., van Rijswijk-Deij, R., Sadre, R., Pras, A.:
  {Measuring the Adoption of DDoS Protection Services}. In: ACM IMC (2016)

\bibitem{BlackholingHoneypots}
Jonker, M., Pras, A., Dainotti, A., Sperotto, A.: {A First Joint Look at DoS
  Atacks and BGP Blackholing in the Wild}. In: ACM IMC (2018)

\bibitem{karami2013booters}
Karami, M., McCoy, D.: {Rent to Pwn: Analyzing Commodity Booter DDoS Services}
  \textbf{38}(6) (Dec 2013)

\bibitem{booterIMC}
Kopp, D., Wichtlhuber, M., Poese, I., de~Santanna, J.J.C., Hohlfeld, O.,
  Dietzel, C.: {DDoS Hide \& Seek: On the Effectiveness of a Booter Services
  Takedown}. In: ACM IMC (2019)

\bibitem{kramer2015raid}
Kr{\"a}mer, L., Krupp, J., Makita, D., Nishizoe, T., Koide, T., Yoshioka, K.,
  Rossow, C.: {AmpPot: Monitoring and Defending Against Amplification DDoS
  Attacks}. In: International Workshop on Recent Advances in Intrusion
  Detection (2015)

\bibitem{2016-Mirai-attack}
Krebs, B.: {KrebsOnSecurity Hit With Record DDoS}.
  \url{https://krebsonsecurity.com/2016/09/krebsonsecurity-hit-with-record-ddos}
  (2016)

\bibitem{IMC04}
Lakhina, A., Crovella, M., Diot, C.: Characterization of network-wide
  anomaliesin traffic flows. In: ACM IMC (2004)

\bibitem{lichtblau17spoofing}
Lichtblau, F., Streibelt, F., Kr{\"u}ger, T., Richter, P., Feldmann, A.:
  {Detection, Classification, and Analysis of Inter-domain Traffic with Spoofed
  Source IP Addresses}. In: ACM IMC (2017)

\bibitem{Port0}
Luchs, M., Doerr, C.: The curious case of port 0. In: IFIP Networking (2019)

\bibitem{spoofer2019}
Luckie, M., Beverly, R., Koga, R., Keys, K., Kroll, J.A., claffy, k.: Network
  hygiene, incentives, and regulation: Deployment of source address validation
  in the internet. In: ACM SIGSAC Conference on Computer and Communications
  Security (2019)

\bibitem{OliPort0}
Maghsoudlou, A., Gasser, O., Feldmann, A.: Zeroing in on port 0 traffic in the
  wild. In: PAM (2021)

\bibitem{bank-attack}
Mohamed, J.: {Daily Mirror: Hackers attack the Stock Exchange: Cyber criminals
  take down website for more than two hours as part of protest against world's
  banks}.
  \url{http://www.dailymail.co.uk/news/article-3625656/Hackers-attack-Stock-Exchange-Cyber-criminals-website-two-hours-protest-against-world-s-banks.html}
  (2016)

\bibitem{Moore2001}
Moore, D., Voelker, G., Savage, S.: {Inferring Internet Denial-of-Service
  Activity}. In: USENIX Security Symposium (2001)

\bibitem{morales17tbps18}
Morales, C.: {NETSCOUT Arbor Confirms 1.7 Tbps DDoS Attack; The Terabit Attack
  Era Is Upon Us}.
  \url{https://asert.arbornetworks.com/netscout-arbor-confirms-1-7-tbps-ddos-attack-terabit-attack-era-upon-us/}
  (2018)

\bibitem{Scrubber}
{Moura}, G.C.M., {Hesselman}, C., {Schaapman}, G., {Boerman}, N., {de Weerdt},
  O.: {Into the DDoS maelstrom: a longitudinal study of a scrubbing service}.
  In: European Symposium on Security and Privacy Workshops (2020)

\bibitem{mskix15}
MSK-IX: {Protection against DDoS-attacks by blackholing},
  \url{www.msk-ix.ru/eng/routeserver.html\#blackhole}

\bibitem{downTheBlackhole}
Nawrocki, M., Blendin, J., Dietzel, C., Schmidt, T.C., Wählisch, M.: {Down the
  Black Hole: Dismantling Operational Practices of BGP Blackholing at IXPs}.
  In: ACM IMC (2019)

\bibitem{netix15}
NETIX: {Blackholing}, \url{www.netix.net/services/14/NetIX-Blackholing}

\bibitem{netscout2020}
Netscout: {Netscout Threat Intelligence Report (2020-02)}.
  \url{https://www.netscout.com/sites/default/files/2020-02/SECR_001_EN-2001_Web.pdf}
  (2020)

\bibitem{nokia-acl}
NOKIA: {Filter Policies}.
  \url{https://documentation.nokia.com/html/0_add-h-f/93-0073-HTML/7750_SR_OS_Router_Configuration_Guide/filters.html}
  (2020), accessed: 2020-05-24

\bibitem{FreeBuff}
null001: {OpenVPN service is used for UDP reflection amplification DDoS
  attack}. \url{http://13.58.107.157/archives/8190} (Sep Sep)

\bibitem{princesmallddos13}
Prince, M.: {The DDoS That Knocked Spamhaus Offline (And How We Mitigated It)}.
  \url{https://blog.cloudflare.com/the-ddos-that-knocked-spamhaus-offline-and-ho/}
  (2013)

\bibitem{princeddos14}
Prince, M.: {Technical Details Behind a 400Gbps NTP Amplification DDoS Attack}.
  \url{https://blog.cloudflare.com/technical-details-behind-a-400gbps-ntp-amplification-ddos-attack/}
  (2014)

\bibitem{RFC900}
Reynolds, J., Postel, J.: Assigned numbers. https://tools.ietf.org/html/rfc900
  (1984)

\bibitem{rossow14amplification}
Rossow, C.: {Amplification Hell: Revisiting Network Protocols for DDoS Abuse}.
  {NDSS}  (2014)

\bibitem{ryba15amplification}
Ryba, F.J., Orlinski, M., W{\"a}hlisch, M., Rossow, C., Schmidt, T.C.:
  {Amplification and DRDoS Attack Defense--A Survey and New Perspectives}.
  arXiv preprint arXiv:1505.07892  (2015)

\bibitem{sachdeva2009performance}
Sachdeva, M., Kumar, K., Singh, G., Singh, K.: {Performance Analysis of Web
  Service Under DDoS attacks}. In: IEEE International Advance Computing
  Conference (2009)

\bibitem{8586810}
{Singh}, K., {Singh}, A.: Memcached ddos exploits: Operations, vulnerabilities,
  preventions and mitigations. In: International Conference on Computing,
  Communication and Security (2018)

\bibitem{akamaiDDoSreport}
Technologies, A.: {2018 State of the Internet / Security: A Year in Review}.
  \url{https://www.akamai.com/us/en/multimedia/documents/state-of-the-internet/2018-state-of-the-internet-security-a-year-in-review.pdf}
  (2018)

\bibitem{thomas2017ecrime}
Thomas, D.R., Clayton, R., Beresford, A.R.: {1000 days of UDP amplification
  DDoS attacks}. In: APWG Symposium on Electronic Crime Research (2017)

\bibitem{ransomware}
Times, N.Y.: {Hackers Hit Dozens of Countries Exploiting Stolen N.S.A. Tool}.
  \url{https://www.nytimes.com/2017/05/12/world/europe/uk-national-health-service-cyberattack.html}
  (2017)

\bibitem{BrianFragmentation}
Trammel, B.: Private conversation (2021)

\bibitem{estonia2019misc}
Traynor, I.: {Russia accused of unleashing cyberwar to disable Estonia}.
  \url{https://www.theguardian.com/world/2007/may/17/topstories3.russia} (2007)

\bibitem{uscert18amplification}
{US-CERT}: {UDP-Based Amplification Attacks}.
  \url{https://www.us-cert.gov/ncas/alerts/TA14-017A} (2018)

\bibitem{vissers15dns-dps}
Vissers, T., Goethem, T.V., Joosen, W., Nikiforakis, N.: {Maneuvering around
  Clouds: Bypassing cloud-based Security Providers}. {ACM CCS}  (2015)

\bibitem{vissers2014ddos}
Vissers, T., Somasundaram, T.S., Pieters, L., Govindarajan, K., Hellinckx, P.:
  {DDoS Defense System for Web Services in a Cloud Environment}. Future
  Generation Computer Systems  \textbf{37},  37--45 (2014)

\bibitem{github}
{ZDNet}: {GitHub hit with the largest DDoS attack ever seen}.
  \url{https://www.zdnet.com/article/github-was-hit-with-the-largest-ddos-attack-ever-seen/}
  (2018)

\bibitem{memcacheZdnet}
ZDNet: Memcached ddos: The biggest, baddest denial of service attacker yet.
  \url{https://www.zdnet.com/article/memcached-ddos-the-biggest-baddest-denial-of-service-attacker-yet/}
  (2018)

\end{thebibliography}

\end{document}